\documentclass[10pt, conference]{IEEEtran}
\IEEEoverridecommandlockouts
\usepackage{cite}
\usepackage{amsmath,amssymb,amsfonts}
\usepackage{algorithmic}
\usepackage{graphicx}
\usepackage{textcomp}
\usepackage{xcolor}
\ifCLASSOPTIONcompsoc           \usepackage[caption=false,font=normalsize,labelfont=sf,textfont=sf]{subfig}
\else
\usepackage[caption=false,font=footnotesize]{subfig}
\fi

\def\BibTeX{{\rm B\kern-.05em{\sc i\kern-.025em b}\kern-.08em
    T\kern-.1667em\lower.7ex\hbox{E}\kern-.125emX}}
\usepackage{romannum}
\usepackage{url}

\begin{document}

\title{A Cooperative Optimal Mining Model for Bitcoin\\}

\author{
\IEEEauthorblockN{David Lajeunesse\IEEEauthorrefmark{1} and Hugo D. Scolnik\IEEEauthorrefmark{2}}
\IEEEauthorblockA{
\IEEEauthorrefmark{1}\IEEEauthorrefmark{2}\textit{Master in Information Security}
\IEEEauthorblockA{
\IEEEauthorrefmark{2}\textit{Department of Computer Science and ICC
} \\}
University of Buenos Aires, Argentina \\
\IEEEauthorrefmark{1}davidlajeunesse1, \IEEEauthorrefmark{2}hscolnik@gmail.com}
}

\IEEEoverridecommandlockouts
\IEEEpubid{\makebox[\columnwidth]{978-1-6654-3924-4/21/\$31.00~\copyright2021 IEEE \hfill} \hspace{\columnsep}\makebox[\columnwidth]{ }}

\maketitle

\IEEEpubidadjcol

\begin{abstract}
We analyze Bitcoin mining from the perspective of a game and propose an optimal mining model that maximizes profits of pools and miners. The model is a two-stage Stackelberg game in which each stage forms a sub-game. In stage \Romannum{1}, pools are the leaders who assign a computing power to be consumed by miners. In stage \Romannum{2}, miners decide of their power consumption and distribution. They find themselves in a social dilemma in which they must choose between mining in solo, therefore prioritizing their individual preferences, and participating in a pool for the collective interest. The model relies on a pool protocol based on a simulated game in which the miners compete for the reward won by the pool. The solutions for the stage~\Romannum{1} sub-game and the simulated protocol game are unique and stable Nash equilibriums while the stage \Romannum{2} sub-game leads to a stable cooperative equilibrium only when miners choose their strategies according to certain criteria. We conclude that the cooperative optimal mining model has the potential to favor Bitcoin decentralization and stability. Mainly, the social dilemma faced by miners together with the balance of incentives ensure a certain distribution of the network computing power between pools and solo miners, while equilibriums in the game solutions provide stability to the system.
\end{abstract}


\section{Introduction}

In Bitcoin, miners are responsible for the transactions added to the chain. Therefore, their role in the system is pivotal as they determine the owners of the transferred funds. Their honest behavior is fostered through financial incentives: they receive a fixed reward for every block extending the main chain and also collect fees associated with the transactions processed. Miners can mine as lone nodes or participate in a mining pool in which they combine their resources to solve the Proof-of-Work (PoW). A solo miner is less likely to solve the PoW first and consequently his income is less constant. On the contrary, in a pool, rewards are more frequent but they must be shared among participating miners.

Currently, in the Bitcoin network, around 65\% of the computing power is controlled by three Chinese pools~\cite{coinshares}. This undermines the principle of decentralization on which relies the security of the system and makes Bitcoin vulnerable to majority attacks~\cite{saad2019exploring} that can be used as levers to alter the transaction history. Decentralization of the computational work is hard to ensure when miners mainly seek to satisfy their personal interests and base their decisions on maximizing their profits using the information at their disposal. In this regard, the overall transparency of mining-related data is lacking as participants have no incentive to share their information. In consequence, there are not enough elements to base an optimal economic decision. Ultimately, miners often favor the income stability offered by pools. A large proportion of miners participate to a limited number of pools whose modalities of operation meet their common needs. As a result, the current mining architecture is insufficient to ensure decentralization and stability of the system as no robust method supports these principles beyond the PoW. This work models Bitcoin mining with game theory. The solution of the game leads to a stable and decentralized cooperative equilibrium for the benefit of all: pools, miners, users and investors.

The rest of the paper is organized as follows. 
Section~\ref{Related Work} summarizes related work that serves as a basis for the optimal mining model proposed in section~\ref{Bitcoin Optimal Mining Model} and its application in a broader model based on cooperation in section~\ref{Bitcoin Cooperative Optimal Mining Model}. Section~\ref{Simulation} shows the results of a simulation, section~\ref{Discussion} discusses and evaluates the cooperative optimal mining model and finally, section~\ref{Conclusion} presents a breve conclusion of this work.

\section{Related Work}\label{Related Work} 

\subsection{Optimal Mining}

Dimitri~\cite{dimitri2017bitcoin} and Chiu et Koeppl~\cite{chiu2019incentive} apply game theory to analyze the optimal power consumption of a mining node. They assume that miners strive to maximize their utilities, i.e., profits, namely revenue minus cost. A similar model has also been developed by Xiong et al.~\cite{xiong2018optimal} in an edge computing service context in support of a blockchain application for mobile devices. They call \textit{miners' demand game} the noncooperative competition between miners for the reward. All three analyzes lead to equivalent results. However, Xiong et al.~\cite{xiong2018optimal} consider more parameters: among others, the transaction fees and the probability of mining an orphan block.

All the studies mentioned above apply a static model to represent the competition between miners: there is only one permanent state of the game that results from all the miners' simultaneous decisions regarding their resource allocation. Equilibrium is achieved when everyone adopts the equilibrium strategy, i.e., optimal strategy. Otherwise, at least one miner can improve his utility by unilaterally changing his strategy. In reality, Bitcoin is an open network and any miner is free to connect, disconnect and modify his mining-allocated resources at any time. The system state and the total computing power are hence inherently dynamic. Consequently, a static model fails to take into account the evolving nature of power consumption resulting from the strategic interactions between miners. 

Dhamal et al.~\cite{dhamal2019stochastic} propose a model to evaluate the optimal power consumption in the dynamic context of players involved in the mining competition. They model a stochastic game that ends when a block is successfully mined. Players are the miners who seek to maximize their utilities and can modify their investment strategies of computing power according to the state of the game and the strategies of others. A game state corresponds to a set of miners present in the system, which they can join and leave at any time. Therefore, unlike the static model, theirs considers the possibility that the reward may not be earned in the current state with the current set of strategies of the miners involved. The transition between states is probabilistic and continuous in time. At each transition, miners reevaluate their strategies and a new set of strategies is defined for the next state. Two state transitions are defined: going from state $S$ to state $S'$ when a new miner enters the system or going from state $S$ to state $S''$ when one leaves. The expected utility of a miner in state $S$ is a weighting of his expected utility if the reward is won in the current state before any transition, his expected utility if the system transits to state $S'$ and his expected utility if it goes to state $S''$. When a transition occurs, the new state becomes the current one, that is $S$, and the expected utility is re-evaluated in the same way. Therefore, the expected utility function of a miner is recursive, which facilitates its optimization in order to evaluate the best response function. By summing all the players, the authors then determine the Markov perfect equilibrium (MPE) that corresponds to the equilibrium strategy.

In their work, Dhamal et al.~\cite{dhamal2019stochastic} present a general stochastic model that allows the analysis of investment strategies of computing power in a distributed system environment. They analyze two scenarios, the first being a generalization of the second. In scenario 1, similar to blockchain mining, a reward is offered for solving a problem. In scenario 2, a reward is offered to participants based on their individual contributions to the total computing power, as in the case of a volunteer computing system for example. In the first scenario, the authors consider that the problem resolution rate varies depending on the current network computing power, while in the second scenario the resolution rate is fixed and independent of the network power. Despite the fact that the Bitcoin protocol adjusts the PoW resolution rate every 2016 blocks according to the time it took to generate them, and therefore according to the network computing power, scenario 2 is better suited for Bitcoin. The reason is that in the closed framework of a game, that is to say the mining of a block, the constant rate of resolution of the PoW is independent of the total network power.

\subsection{Cooperation in Social Dilemmas} \label{Cooperation in Social Dilemma}

Let us consider a game with $n$ players in which possible actions are cooperation ($C$) and desertion ($D$). We denote the utility of a cooperating player $a_j$ and that of a deserting player $b_j$, where $j$ is the number of others players who cooperate. We have the following properties of the utilities of a social dilemma~\cite{kC}:
\begin{enumerate}
\item Players prefer others to cooperate: $a_{j+1}\geqslant a_j$ and $b_{j+1}\geqslant b_j$, $\forall j$
\item Desertion generates an expected utility strictly superior to cooperation: $b_{j+1}>a_j$, $\forall j$  
\item Mutual cooperation is promoted at the expense of mutual desertion: $a_{n-1}>b_0$
\end{enumerate}

Hilbe et al.~\cite{hilbe2014cooperation} developed a theory for Zero-Determinant (ZD) strategies (Press et Dyson~\cite{press2012iterated}) that allows to achieve cooperative equilibriums in repeated multiplayer social dilemmas, unlimited in the number of players\footnote{For large groups, the mechanisms developed to maintain equilibrium are often ineffective because it is difficult for players to analyze the behaviors of their opponents and have an influence over them. Therefore, the theory of Hilbe et al.~\cite{hilbe2014cooperation} fits well with the high number of Bitcoin miners.}. ZD strategies are \textit{memory-one}, which means they are based solely on the result of the previous iteration of the game. From Akin's~\cite{akin2016iterated} overall result\footnote{$(\boldsymbol{p}-\boldsymbol{p}^{Rep})\cdot \boldsymbol{v}=0$, where $\boldsymbol{p}^{Rep}$ is the memory-one \textit{repeat} strategy with $p_{C,j}^{Rep}=1$ and $p_{D,j}^{Rep}=0$ and $\boldsymbol{v}$ is the limit distribution.} showing the relationship between a memory-one strategy and the limit distribution, Hilbe et al.~\cite{hilbe2014cooperation} demonstrate the linear relationship between the utility of a player who applies a ZD strategy of a certain form and that of others, regardless of their strategies. By choosing well the parameter values, a player can thus exercise some control over the relationship between its utilities and those of the rest.

Hilbe et al.~\cite{hilbe2014cooperation} show that certain ZD strategies lead to a stable cooperative Nash equilibrium if they respect two criteria. One of them infers that when the number of players is large, they can always achieve such equilibrium if they adopt a \textit{fair} ZD strategy of the form $\boldsymbol{p}=\boldsymbol{p}^{Rep} +\phi [\boldsymbol{g}^i-\boldsymbol{g}^{-i}]$. The vector $\boldsymbol{p}=(p_{C,n-1},\dots,p_{C,0},p_{D,n-1},\dots,p_{D,0})$ contains the probabilities $p_{S,j}$ that player $i$ will cooperate in the next iteration of the game having chosen action $S\in{(C,D)}$ in the previous one while $j$ others cooperated. The vector $\boldsymbol{g}^i=(g_{S,j}^i)$ comprises the possible utilities for player $i$ during an iteration, and we denote $g_{C,j}^i=a_j$ and $g_{D,j}^i=b_j$. Similarly, $\boldsymbol{g}^{-i}=(g_{S,j}^{-i})$ corresponds to the average utilities of others, with $g_{C,j}^{-i}=\frac{ja_j+(n-j-1)b_{j+1}}{n-1}$ and $g_{D,j}^{-i}=\frac{ja_{j-1}+(n-j-1)b_j}{n-1}$. The parameter $\phi$ determines the speed, in terms of number of iterations of the game, at which utilities converge towards the linear relationship. As the probabilities of cooperating $p_{S,j}$ are necessarily between 0 and 1 inclusively, the value of $\phi$ must be chosen accordingly and the permissible values depend on the type of social dilemma. By applying a fair strategy, a player ensures that he will have a utility equal to the average utility of others, regardless of their strategies. If enough players adopt such a strategy, they can maintain a stable cooperative equilibrium and prevent mutual desertion caused by a recalcitrant group or in the event of accidental desertion.

\section{Bitcoin Optimal Mining Model}\label{Bitcoin Optimal Mining Model}

We use the stochastic model of Dhamal et al.~\cite{dhamal2019stochastic} as the basis, to which we integrate some parameters of the static model of Xiong et al.~\cite{xiong2018optimal} to consider the transaction fees and the probability of a mined block ending up orphaned. We introduce an energy efficiency factor $k$ to properly model the probability of wining the reward based on the number of PoW computed solutions instead of the energy consumed. The value of $k$ depends on the mining hardware and can be taken from the product characteristics. For example, the energy efficiency of an Antminer S19 Pro is approximately $1/30$ TH/J~\cite{Bitmain}. The criterion to fulfill is the relative value of the $k_i$ factors among players and not the value in itself. Finally, we break down the value of the reward by including a currency conversion rate $\tau$ to isolate the impact of the value of the Bitcoin currency on mining. The rest of the notation is mostly consistent with~\cite{dhamal2019stochastic}, as shown in Table \ref{tab1}.

\begin{table}[!b]
\caption{Notation of the Bitcoin optimal mining model}
\begin{center}
\begin{tabular}{|c|p{7cm}|}
\hline 
$\mathcal{U}$ & Universal set of strategic players, present in the system or not\\
\hline
$S$ & Set of strategic players currently present in the system\\
\hline
$\hat{S}$ & Set of players in $S$ who invest a positive computing power\\
\hline
$x_i^{(S)}$ & Computing power investment strategy of player $i$ in state $S$\\
\hline
$l$ & Constant power consumption of the nonstrategic players\\
\hline
$\lambda_i$ & Arrival rate in the system of strategic player $i$\\
\hline
$\mu_i$ & Departure rate from the system of strategic player $i$\\
\hline
$\beta$ & Constant rate of resolution of the PoW per unit of time\\
\hline
$\frac{1}{D^{(S,x)}}$ $^{\mathrm{a}}$ & Sojourn time in state $S$\\
\hline
$r$ & Fixed block reward in bitcoins\\
\hline
$\theta$ & Average fees of a transaction in a block of player $i$\\
\hline
$\tau$ & Bitcoin currency conversion rate\\
\hline
$t_i$ & Number of transactions included in a block of player $i$\\
\hline
$z_i$ & Propagation delay factor of a block in the network\\
\hline
$c_i$ & Marginal cost incurred by player $i$ per power and time unit\\
\hline
$k_i$ & Energy efficiency factor of player $i$\\
\hline
$k_l$ & Average energy efficiency factor of the nonstrategic players\\
\hline
\multicolumn{2}{l}{$^{\mathrm{a}}$$D^{(S,x)}=\beta +
\sum_{j\notin S}{\lambda_j} + \sum_{j\in S}{\mu_j}$, where $D^{(S,x)}$, $\beta$, $\lambda_j$ and ${\mu_j}$}\\
\multicolumn{2}{l}{follow exponential distributions.}
\end{tabular}
\label{tab1}
\end{center}
\end{table}

The expected utility of player $i$ in state $S$ adopting strategy $x$ corresponds to its expected revenue minus its expected cost, as follows:

\begin{multline}
R_i^{(S,x)} = \frac{\beta}{D^{(S,x)}}\frac{k_ix_i^{(S)}}{\sum_{j\in S}{k_jx_j^{(S)}}+k_ll}\tau(r+\theta t_i)e^{-\beta z_i t_i }\\
-\frac{c_i x_i^{(S)}}{D^{(S,x)}}+\sum_{j\notin S}{\frac{\lambda_j}{D^{(S,x)}}R_i^{(S\cup{\{j\}},x)}}\\
+\sum_{j\in S}{\frac{\mu_j}{D^{(S,x)}}R_i^{(S\setminus{\{j\}},x)}}
\label{eq1}
\end{multline}

The changes we make to the model of Dhamal et al.~\cite{dhamal2019stochastic} have no impact on their demonstration of the existence and uniqueness of the MPE and it remains valid for our model. Hence, we use the same methodology as in~\cite{dhamal2019stochastic} to determine the MPE that corresponds to the equilibrium strategy, with the following result (proof in appendix):

\begin{equation}
x_i^{(S)^*}=\max 
\bigg\{ 
\psi^{(S)}
\bigg( 
\frac{1}{k_i}-\frac{\psi^{(S)}} {{k_i}^2 \beta\tau(r+\theta t_i) e^{-\beta z_i t_i}} c_i 
\bigg),0 
\bigg\} 
\label{eq2}
\end{equation}
where,
\begin{equation*}
\psi^{(S)}=\frac{|\hat{S}|-1+\sqrt{
{(|\hat{S}|-1)}^2+\frac{4k_l l}{\beta\tau} \sum_{j\in \hat{S}}{\frac{c_j}{k_j (r+\theta t_j) e^{-\beta z_j t_j } }}}}
{\frac{2}{\beta\tau} \sum_{j\in \hat{S}}{\frac{c_j}{k_j (r+\theta t_j) e^{-\beta z_j t_j } }}}
\end{equation*}

Based on \eqref{eq2} and following the same reasoning as in~\cite{dhamal2019stochastic}, a player will not invest any computing power if $ \frac{1}{k_i} \leqslant \frac{\psi^{(S)}}{{k_i}^2 \beta\tau(r+\theta t_i )e^{-\beta z_i t_i} } c_i$. To participate in mining, we then have the condition $\frac{1}{k_i} > \frac{\psi^{(S)}}{{k_i}^2 \beta\tau(r+\theta t_i )e^{-\beta z_i t_i} } c_i$  or equivalently $ \frac{c_i}{k_i} < \frac{\beta\tau(r+\theta t_i )e^{-\beta z_i t_i}}{\psi^{(S)}}$  . Replacing $\psi^{(S)}$, we obtain:

\begin{equation}
\frac{c_i}{k_i}<\frac{(r+\theta t_i) e^{-\beta z_i t_i} \cdot 2\sum_{j\in \hat{S}}{\frac{c_j}{k_j (r+\theta t_j) e^{-\beta z_j t_j}}}}{
|\hat{S}|-1+\sqrt{
{(|\hat{S}|-1)}^2+\frac{4k_l l}{\beta\tau} \sum_{j\in \hat{S}}{\frac{c_j}{k_j (r+\theta t_j) e^{-\beta z_j t_j } }}}}
\label{eq3}
\end{equation}

Thus, there is a maximum threshold for the ratio of the energy cost over the energy efficiency that depends on various parameters and beyond which a player will choose not to mine.

\section{Bitcoin Cooperative Optimal Mining Model}\label{Bitcoin Cooperative Optimal  Mining Model}

\begin{figure}[!b]
\centering{\includegraphics[width=0.5\textwidth]{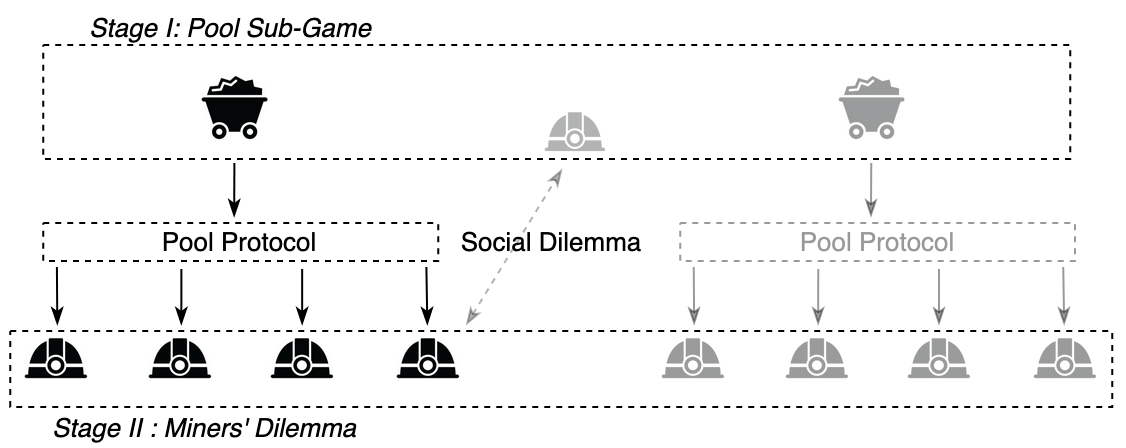}}
\caption{Stackelberg game of the cooperative optimal mining model for Bitcoin.}
\label{fig1}
\end{figure}

The cooperative optimal mining model is a two-stage Stackelberg game. Stage \Romannum{1} is the \textit{pool sub-game}, in which pools are the leaders who work together to determine the optimal computing power that each should generate. Stage \Romannum{2} is the \textit{miners' dilemma}, a sub-game in which miners are the followers who decide on their participation in a pool and their investment strategy of computing power. This is schematized in Fig.~\ref{fig1}. 

In stage \Romannum{1}, pools collaborate together by sharing the information necessary to achieve equilibrium and maximize their utility. The information shared results from the cooperation among miners; by joining a pool, they agree to share their marginal cost, maximum computing capacity and energy efficiency, for the benefit of all. This allows the pool to establish the computational work that each of its members must do to maximize his utility. However, this optimal investment represents a logical course of action for the miners rather than an obligation, as it cannot be imposed. When the pool wins a reward, it is shared among miners based on the outcome of a simulated game, which we call the \textit{protocol game}. The latter, the distribution of the computational work and the sharing of the reward are all components of the pool protocol that governs the rules of operation between the pool and its miners. The miners' dilemma is a noncooperative game between miners in the Bitcoin network, in which they decide on their participation in a pool and their investment of computing power.

\subsection{Pool Sub-Game}
The pool sub-game is the competition between pools for the reward associated with the mining of a block extending the chain. We apply the Bitcoin optimal mining model presented in section \ref{Bitcoin Optimal Mining Model} to analyse the game, where pools are the players. In \eqref{eq1}, we set $\mu_j=\lambda_j=0$ as pools are considered always present in the system and, therefore, we obtain a static model. The equilibrium is given by \eqref{eq2}. As for the other parameters, $S$ is the set of pools present in the system with sufficient resources to play the optimal strategy; $l$ is the residual network computing power that is not generated by $S$; $k_i$ is the energy efficiency of pool $i$ and corresponds to an average of the energy efficiencies of its miners weighted by their contribution to the optimal strategy\footnote{Collaboration between pools enables an iterative calculation to evaluate their marginal cost, energy efficiency and optimal strategy.}; $c_i$ is the marginal cost of pool $i$ and is an average of the marginal costs of its members weighted by their contribution to the optimal strategy; $\theta$ corresponds to the average fee of a transaction included in a block of pool $i$; $z_i$ is the propagation delay factor in the network of a block of pool $i$. We make the assumption that every major pool invests a positive computing power supposing a relatively uniform distribution among miners with different marginal costs and energy efficiencies. Therefore, we define without distinction $\hat{S}=S=\mathcal{U}$.

\subsection{Pool Protocol}\label{Pool Protocol}

\subsubsection{Distribution of the computational work} The pool calculates the computational work that each miner must do. To maximize the pool utility, the distribution of the work is carried out in increasing order of ratios ${c_i}/{k_i}$ and equally among miners, according to their maximum capacity and until the pool equilibrium strategy is reached.

\subsubsection{Protocol game} This component simulates a competition between miners of the pool for the reward that the latter earned. The equilibrium is given by \eqref{eq2} and serves to calculate the reward fraction of each miner. We set $\theta=t_i=z_i=0$ since a block is broadcast in the pool sub-game and only the header is transmitted to the miners to solve the PoW. As for the other parameters, $S$ is the set of miners registered in the pool who have sufficient resources to play the optimal strategy; $k_i$ is the energy efficiency of miner $i$; $l$ is the computing power generated by the miners who fulfill condition \eqref{eq3} but don't have enough resources to be part of set $S$ (those miners invest their maximum capacity); $\lambda_i$ and $\mu_i$ refer to the connection/disconnection rates to/from the pool of miner $i$; $c_i$ is the marginal cost of miner $i$, and $\hat{S}$ is the set of miners in the pool who invest a positive computing power. The latter set is iteratively established by gradually integrating miners in increasing order of ratios ${c_i}/{k_i}$ until the condition \eqref{eq3} is met.

\subsubsection{Sharing of the reward}\label{Sharing of the reward} The reward won by the pool is shared among miners according to the cost of their resources and in proportion to the optimal rate of return (ROI)\footnote{$ROI=Expected~utility/Expected~cost$.} they would have had if they had participated in an intra-pool competition as modeled by the protocol game. The fraction of the pool reward distributed to a miner corresponds to the proportion of his expected reward ($\mathbb{E}[r]_i^{x_i}$) over the expected pool reward (${\mathbb{E}[r]}_p^{x_p^*}$). In each case, the expected reward corresponds to the expected utility plus the expected cost. Hence, the fraction $\alpha_i$ of the reward given to miner $i$ is
$
\alpha_i=\frac{\mathbb{E}[r]_i^{x_i^*}}{\mathbb{E}[r]_p^{x_p^*}}
=\frac{\mathbb{I}_i \cdot R_p^{x_p^*} + \frac{c_i x_i^*}{\beta}}{R_p^{x_p^*} + \frac{c_p x_p^*}{\beta}}
$, 
where $\mathbb{I}_i=\frac{I_i^C \cdot I_i^{ROI}}{\sum_{j \in P}{I_j^C \cdot I_j^{ROI}}}$. We denote $I_i^C$ the relative cost\footnote{A parameter relative rate for a certain group is the ratio of the parameter value for that group over that of an arbitrary chosen reference group.} of the resources consumed by miner $i$; $I_i^{ROI}$ is miner $i$ expected relative ROI in the protocol game; $P$ is the set of miners in the pool; $R_p^{x_p^*}$ is the pool expected utility calculated in the pool sub-game; $x_i^*$ is the optimal computational work for miner $i$, and $x_p^*$ is the pool equilibrium strategy in the pool sub-game. Relative costs are used to equitably share the reward among miners considering that not all invest the same resources as determined by the distribution method of the computational work. Relative ROIs are calculated from miners’ expected utilities in the protocol game when they adopt their optimal strategy and ensure that they obtain an optimal rate of return compared to other pool members. This deterministic method of sharing the reward based on a simulated game avoids real competition between miners and the consequent reduction of their expected profits. Together, the methods for distributing the computational work and sharing the reward minimize the cost of the pool's optimal strategy and maximize its expected utility by first allocating work to the most efficient miners, i.e., those with lower ${c_i}/{k_i}$ ratios, without affecting their relative optimal ROI.

\subsection{Miners' Dilemma}

The strategy of a miner is to distribute his resources between pools and solo mining. Based on the evaluation of a single block, the expected utility $R_i^x$ of miner $i$ adopting strategy $x$ corresponds to the sum of his expected utilities in pools and as a solo miner, and can be expressed as $R_i^x=\sum_{j \in S}{\Big(\mathbb{I}_{i,p_j}\cdot R_{p_j}^{x_{p_j}}\Big)}+R_{i,m}^x$  where $\mathbb{I}_{i,p_j}$ is the relative rate $\mathbb{I}_{i}$, as defined in section \ref{Sharing of the reward}, of miner $i$ in pool $p_j$; $R_{p_j}^{x_{p_j}}$ is the expected utility of pool $p_j$ adopting strategy $x_{p_j}$ in state $S$ in the pool sub-game; $R_{i,m}^x$ is the expected utility of miner $i$ with strategy $x$ in state $S$ as a solo miner in the pool sub-game, in which he is part of set $S$ if he has sufficient resources to adopt the equilibrium strategy or part of $l$ otherwise.

For a given pool and in the evaluation of a single iteration of the game, i.e., a mined block, a miner will choose to mine in solo if his expected utility is greater than that in a pool, that is if $R_i^x>\mathbb{I}_{i,{p_j}}\cdot R_{p_j}^{x_{p_j}}$. However, this condition is always met with the structure established by the cooperative optimal mining model and is not sufficient for a miner to make a strategic decision that maximizes his utility over several iterations. Pools are alliances of miners and allow to analyze the pool sub-game based on the pools' actions rather than those of their members, who may have individual preferences that differ from those of the group. In the miners' dilemma, this divergence of preferences is at the heart of their strategic choice. Alliances in the pool sub-game reduce competition and increase expected profits. When all the miners cooperate by participating in a pool, their expected utility is greater than that of a noncooperative scenario. However, one who deserts for mining in solo increases his expected utility by benefiting from the reduced competition in the pool sub-game without having to share his rewards. Thus, based on individual rationality, the optimal strategy for a miner is always to desert. However, if everyone adopts this strategy, cooperation disappears and the expected utility of everyone decreases because of the high competition in the pool sub-game. In game theory, this type of problem constitutes a \textit{social dilemma} and finds its name in the dilemma generated for players, who must choose between their individual interest and the collective interest. 

To properly assess the miners’ dilemma, we need to model an endlessly repeating game in which the expected utility of a miner is the sum of his expected utilities over all future iterations. Possible actions are cooperation (C), i.e., participation in a pool, or desertion (D), i.e., mining in solo. Hence, there are $2^n$ possible states of an iteration according to the choices of the $n$ miners present in the system. If the latter have the same utility structure, the space is then limited to $n$ states. As the game is repeated endlessly, players cannot use backward induction\footnote{In game theory, backward induction is a common concept of solution which is based on the actions taken during the last iteration.} to determine their strategy. Therefore, they often rely on the past behavior of their opponents\footnote{This branch of game theory is called evolutionary game theory (EGT), which applies to the study of changing populations~\cite{sandholm2020evolutionary}.}. Accordingly, we rely on the theory of Hilbe et al.~\cite{hilbe2014cooperation} and state that miners adopt a fair ZD strategy of the form $\boldsymbol{p}=\boldsymbol{p}^{Rep} +\phi [\boldsymbol{g}^i-\boldsymbol{g}^{-i}]$ to reach a stable cooperative equilibrium.

When a miner chooses cooperation, his optimal strategy is to do the computational work assigned by the pool. When deserting for solo mining, he integrates the pool sub-game. If he has sufficient resources, he adopts the equilibrium strategy and is a strategic player of the set $S$, otherwise, his power consumption is counted in $l$. In this case, he will invest his maximum capacity if he fulfills condition \eqref{eq3}, the calculation of which he must estimate as he does not share information with the pools as a result of being a nonstrategic player. 

\section{Simulation}\label{Simulation}

\subsection{Pool Sub-Game}\label{Simulation Pool sub-game}

We set the following values for the parameters: $r=6.25$; $\tau=10,000$; $\beta=0.1$; $\theta=0.00012$; $t_i=t_j=2100$; $|\mathcal{U}|=|S|=|\hat{S}|=10$; $c_i=c_j=0.0007$; $\lambda_j=\mu_j=0$; $k_i=k_j=k_l=1$; $z_i=z_j=0.005/60$ and $l=700,000$. Based on \eqref{eq2}, we find that with ten strategic pools present in the system, the optimal strategy is to invest $759,174.7$ kWh/min, which generates an expected profit of $\$535.60$ per block for each strategic pool as determined by \eqref{eq1}.

\subsection{Pool Protocol}

To facilitate calculations, we suppose all miners have the same marginal cost and energy efficiency. We suppose that $1100$ miners are registered in each pool and $1050$ are connected. We set that $500$ other miners have a maximum computing capacity that impedes adopting the equilibrium strategy and, therefore, their maximum computing power is counted in $l$. Hence, we set the following parameters for the protocol game, where $\mathbb{E}[r]_p^{x_p^*}$ is the expected reward earned by the pool when adopting the optimal strategy $x_p^*$ in the pool sub-game: $r=\mathbb{E}[r]_p^{x_p^*}=5849.82$; $\tau=1$; $\beta=0.1$; $|\mathcal{U}|=600$; $|S|=|\hat{S}|=550$; $c_i=0.0007$; $l=10,000$; $\lambda_j=1$; $\mu_j=0.1$; $k_i=k_j=k_l=1$, and the other parameters are null.

From the pool optimal power investment calculated in section \ref{Simulation Pool sub-game}, we establish the distribution of the computational work between miners and derive the work cost per miner and the relative cost ratios. After, we evaluate the expected utility of strategic miners in the protocol game. We do the same for nonstrategic miners, using the same method as for strategic ones. However, we use a constant power investment corresponding to their maximum capacity, which is assumed to be the same for all, that is $x_i=l/500$. Expected utilities in the protocol game only serve to determine the relative ROIs, which are used to calculate the reward fraction $\alpha_i$ and the real expected utility of each miner. All the calculations are performed as described in section~\ref{Pool Protocol}. 

Table \ref{tab2} presents the relation between miners' expected utilities according to three scenarios: mutual cooperation of all the miners in the ten pools ($a_{n-1}$), unilateral desertion of one miner in a pool ($a_{n-2}$ and $b_{n-1}$) and mutual desertion of all ($b_0$). The results obtained respect the properties of the utilities in a social dilemma and show that cooperation significantly increases expected utilities.

\begin{table*}[!t]
\caption{Miners' expected utilities according to different strategic profiles and scenarios of cooperation}
\begin{center}
\begin{tabular}{|c|c|c|c|c|c|c|c|}
\hline
\textbf{Type of}&\textbf{Number of}&\textbf{Maximum Capacity}&\textbf{Assigned Work}&\multicolumn{2}{|c|}{\textbf{Social Dilemma Property 1 (SDP1)}}&\textbf{SDP2}&\textbf{SDP3}\\
\cline{5-8}
\textbf{Miners}&\textbf{Miners}&\textbf{(kWh/min)}&\textbf{(kWh/min)}&\textbf{$\boldsymbol{a_{n-1}/a_{n-2}}$}&\textbf{$\boldsymbol{b_{n}/b_{n-1}}$}&\textbf{$\boldsymbol{b_{n}/a_{n-1}}$}&\textbf{$\boldsymbol{a_{n-1}/b_0}$}\\
\hline
NonStrategic&500&$20$&$20$&$1.000008272$&$1.000026223$&$1.00015$&$596.33$\\
\hline
\cline{2-8}
&300&$2000$&$1362$&$1.000530856$&$1.000287113$&$1.47$&$526.89$\\
\cline{2-8}
Strategic&200&$3000$&$1362$&$1.000793597$&$1.000431086$&$2.20$&$526.89$\\
\cline{2-8}
&50&$5000$&$1362$&$1.001319688$&$1.000718589$&$3.67$&$526.89$\\
\hline
\end{tabular}
\label{tab2}
\end{center}
\end{table*}

\subsection{Miners' Dilemma}

A Markov chain Monte Carlo simulation is used to verify that a stable cooperative Nash equilibrium is achievable. To simplify, we suppose a parametric network of $10,000$ miners, all with the same profile. The vector of the possible utilities for the mining of a block is defined $\boldsymbol{g}^i=(a_{n-1},..., a_0,b_{n-1},...,b_0)$ and we arbitrarily establish $a_{n-1}=\$35.00$, $a_0=\$0.04$, $b_{n-1}=\$70.00$ and $b_0=\$0.05$. We determine the utilities $a_j$ and $b_j$, where $j\in \{1,...,n-2\}$, in order to obtain a linear decrease. The resulting vector $\boldsymbol{g}^i$ respects the properties of the utilities of a social dilemma and $\boldsymbol{g}^{-i}$ is calculated according to the formula given in section~\ref{Cooperation in Social Dilemma}. All miners in the network are assumed to adopt a fair ZD strategy as determined by Hilbe et al.~\cite{hilbe2014cooperation} to achieve a stable cooperative equilibrium.

\begin{figure}[!b]
\centering
\subfloat[]{\includegraphics[width=.5\linewidth]{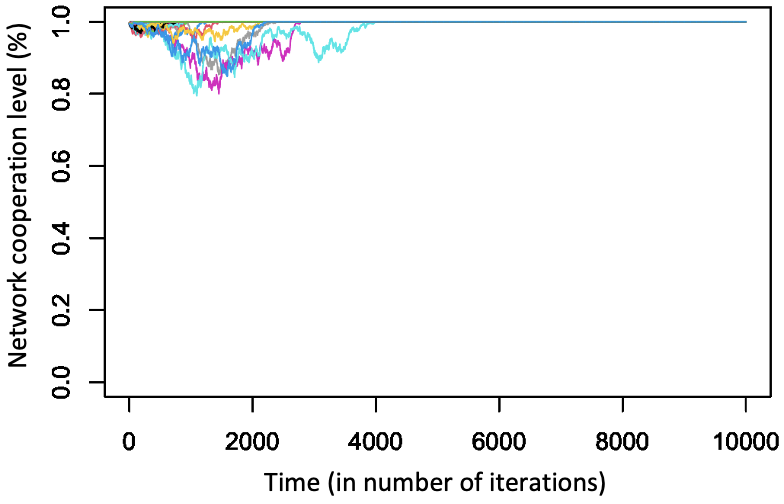} 
\label{fig2:subfig1}}
\subfloat[]{\includegraphics[width=.5\linewidth]{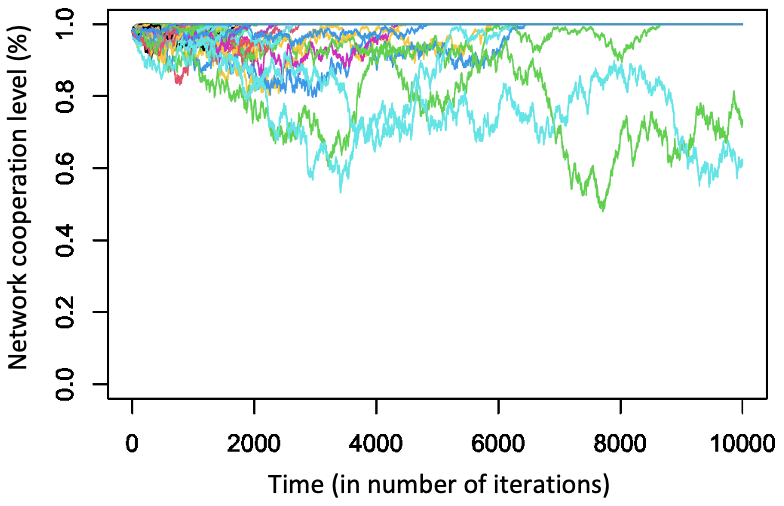} 
\label{fig2:subfig2}}
\vfil
\subfloat[]{\includegraphics[width=.5\linewidth]{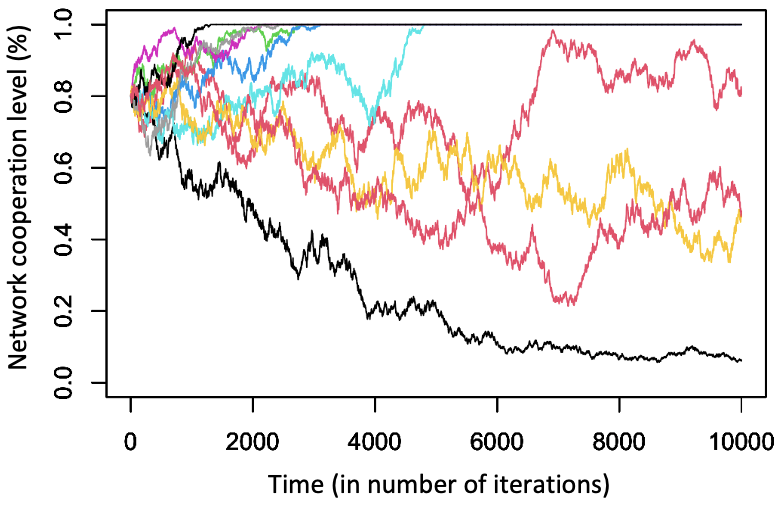} 
\label{fig2:subfig3}}
\subfloat[]{\includegraphics[width=.5\linewidth]{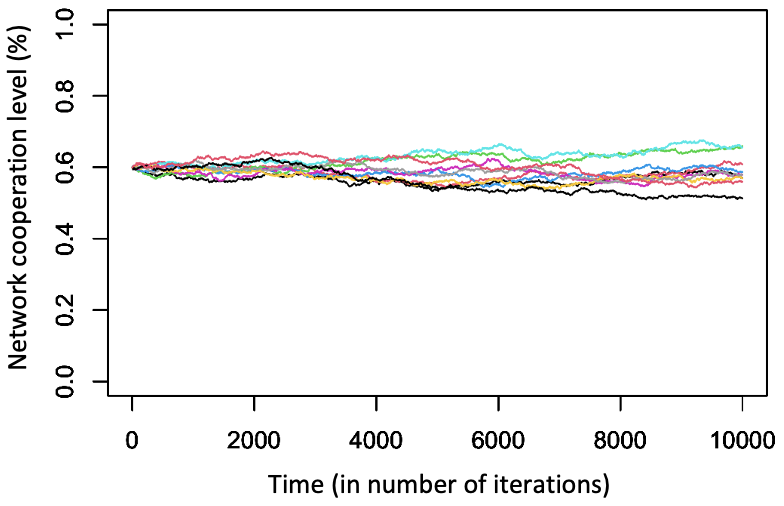} 
\label{fig2:subfig4}}
\caption{Evolution of the cooperation in the network according to different values of $\phi$ and degrees of initial cooperation (IC). (a) 100 simul., $\phi=1/\max(\boldsymbol{g}^i-\boldsymbol{g}^{-i})$, IC = $99.5\%$. (b) 100 simul., $\phi=1/\max(\boldsymbol{g}^i-\boldsymbol{g}^{-i})$, IC = $98\%$. (c) 10 simul., $\phi=1/\max(\boldsymbol{g}^i-\boldsymbol{g}^{-i})$, IC = $80\%$. (d) 10 simul., $\phi=1/[100*\max(\boldsymbol{g}^i-\boldsymbol{g}^{-i})]$, IC = $60\%$.} 
\label{fig2}
\end{figure}

Fig. \ref{fig2} shows the evolution of the degree of cooperation in the network according to different values of $\phi$ and initial cooperation degrees. Fig. \ref{fig2}a,b show that while miners first cooperate at their arrival in the system, a fair strategy leads to a stable cooperative equilibrium and supports to some extent strategy deviations (accidental or voluntary). However, Fig. \ref{fig2}b,c show that for a high $\phi$ value, the more miners desert in an iteration, the harder it will be to strike a stable cooperative equilibrium. This is due to the increased variability in the probabilities of cooperation when $\phi$ is high. On the one hand, desertion of a minority is punished more severely by generating further desertions. On the other hand, the probability of returning to cooperation after deserting decreases significantly with the cooperation degree. As a result, more iterations are needed to achieve a stable equilibrium and the probability for the latter to reach mutual desertion increases as the cooperation decreases. Nevertheless, Fig. \ref{fig2:subfig4} shows that a low $\phi$ value, or even one that tends to zero, stabilizes the current degree of cooperation in the network. For instance, in Fig. \ref{fig2}b,c  where all miners apply a strategy with a large $\phi$, some curves are moving towards mutual desertion and at some point reach a degree of cooperation of $60\%$; cooperative miners could then form an alliance\footnote{Hilbe et al.~\cite{hilbe2014cooperation} also demonstrate the influence that alliances can have on other players. Depending on its size and strategy, an alliance can even incite and bring deserters back to cooperation.} and agree to adjust their $\phi$ to a very low value to ensure a minimum degree of cooperation, regardless of the deserter strategies. The degree of cooperation will thus stabilize as in Fig. \ref{fig2:subfig4} if defectors continue to defect or increase if they return to cooperation.  

\section{Discussion}\label{Discussion}

At equilibrium, the Bitcoin network power consumption obtained from the simulation is of the order of 4000 TWh annually. This is significantly higher than the estimates reported by~\cite{coinshares} and~\cite{Digiconomist}, which are around 60 to 75 TWh annually. This implies that equilibrium would not be reached in the current Bitcoin environment. However, variations in the model parameters could change this situation. In particular, the fixed reward will decrease over time until the transaction fees are the only economic incentive. Moreover, the Bitcoin conversion rate directly affects the reward value.

In Table \ref{tab2}, simulation results also show that the limited power consumption that a pool must generate to reach equilibrium impedes its miners that have additional resources. If they cannot exploit all of their resources within the pool and they don't spend their extra capacity elsewhere, these miners are those making the biggest sacrifice in participating in a pool. Miners will rationally invest their additional available resources in other pools that need them to play the equilibrium strategy or they will mine in solo. It is difficult to estimate the number of players who would be strategic in the pool sub-game. The best scenario, that does not seem unlikely and would ensure decentralization, would be that the implementation of the model creates a competitive market between pools somewhere between an oligopoly and a perfect competition, with a large number of them applying substantially the same protocol. Nonetheless, this aspect should be studied more.

In each sub-game of the cooperative optimal mining model, there is an equilibrium. In the pool sub-game and the protocol game, equilibriums are unique and therefore necessarily stable. In the miners' dilemma, a stable cooperative equilibrium is achievable. In this case, equilibrium does not mean that all miners cooperate at all times, but rather that their strategies are defined to promote cooperation at the expense of desertion. They can establish a positive probability of defecting from the pool under certain circumstances, but will ultimately benefit from going back to cooperation when the desertion rate is high. The stable cooperative equilibrium ensures that there is always a certain degree of cooperation in the network, so that miners who participate in a pool benefit from this cooperation even if there is a recalcitrant group who defects and whose expected utility is greater\footnote{This is aligned with the concept of evolutionary stable strategy (ESS) in symmetric normal form games. A population in which all members play an ESS is resistant to a small recalcitrant group who plays another mixed strategy~\cite{sandholm2020evolutionary}.}. On the one hand, the fact that miners have individual preferences for desertion contributes to decentralization by distributing power outside the pools. On the other hand, the collective benefit from cooperation prompts them to participate in a pool. To this extent, the incentive to cooperate is essential for the viability of the model. Since pools are those who implement it, they must benefit from doing so and be able to attract miners. The stable income offered by a pool is also an incentive to participate in it. In short, there is a balance between incentives for cooperation and desertion that leads to a stable and decentralized cooperative equilibrium of the network power distribution. However, the existence of equilibriums in game solutions does not necessarily mean that they are reached. To do so, players must adopt the equilibrium strategy. This is particularly true in multiplayer social dilemmas that repeat endlessly, since players often define their strategies based on the past behavior of their opponents such that the evolution of the game can be difficult to predict.

Additional pool protocol rules could be implemented. This could include retroactive adjustments to ensure that the rewards earned by miners match those of the equilibrium scenario. Likewise, the same mechanism could be implemented between pools, which would confirm their mutual commitment to respect the equilibrium strategy. This would add an incentive for both miners and pools to adhere to the equilibrium strategy and thus strengthen the stability of the system. The integration of a parameter to incentivize miners to use renewable energy could also be considered. Furthermore, pool fees could correspond to a percentage of the miners’ profits. As the cooperative optimal mining model seems to have the potential to significantly increase miners’ expected utilities, this would imply the same for pool profits.

A mining model based on optimal power consumption offers the advantage to be predictive and brings stability to the system. For example, the model can anticipate the impact of a halving on the network power consumption and thus mitigate the historic uncertainty induced by this operation. Similar analyzes can be conduced from other parameters and serve as a tool for system design decision making. For instance, the effect on mining of modifying the block size limit can be evaluated from the propagation delay $z_i$ and the number of transactions $t_i$ in a block. The model can also help to determine the amount of transaction fees $\theta$ necessary to maintain an incentive for miners and a sufficient network power to ensure the security of the system. As well, the increased transparency of mining information that results from sharing marginal costs, energy efficiencies and maximum capacities facilitates analyzes and forecasting, which in turns strengthen user confidence and system stability.

A deterministic and transparent model regarding power consumption may serve as a preventive or detective tool against certain known mining-related attacks, such as for selfish mining~\cite{sapirshtein2016optimal},~\cite{eyal2014majority} or block retention~\cite{eyal2015miner}. By knowing the resources that each one should consume, it is possible to deduce the rewards that each one should collect and thus detect dishonest behaviors when miners or pools collect more than their fair share. Furthermore, a retroactive adjustments system as suggested above would bring stability and constitute an indirect commitment from pools and miners to act honestly, as it would nullify any extra benefits obtained from dishonest behaviors. However, this should be analyzed in more depth.

\section{Conclusion}\label{Conclusion}

We believe that the proposed model can improve the decentralization, stability and security of Bitcoin. Nevertheless, more complex simulations should be done to evaluate its applicability in the current Bitcoin environment. We extend our conclusions to similar PoW-based blockchains, especially if incentives are designed considering the model.  

Possibly, the implementation could be done through smart contracts. For example, Ethereum layer 2 solutions that allow complex off-chain operations could be considered to constantly evaluate the optimal power consumption of the participants. Oracles could provide external information such as electricity cost, Bitcoin conversion rate and hardware energy efficiency. Furthermore, a decentralized governance system could decide of the evolution of the protocol.

\section*{Appendix}
\setcounter{subsection}{0}

\subsection{Proof of \eqref{eq2}}

\begin{IEEEproof}
Following the same reasoning as in~\cite{dhamal2019stochastic}, optimizing \eqref{eq1} is the same as optimizing 
$Z_i^{(S,x)}=\frac{\beta}{D^{(S,x)}}\frac{k_ix_i^{(S)}}{\sum_{j\in S}{k_jx_j^{(S)}}+k_ll}\tau(r+\theta t_i)e^{-\beta z_i t_i }
-\frac{c_i x_i^{(S)}}{D^{(S,x)}}$. We have $\frac{\partial Z_i^{(S,x)}}{\partial x_i^{(S)}}=\frac{\beta}{D^{(S,x)}}\tau(r+\theta t_i)e^{-\beta z_i t_i }\frac{\partial \alpha_i}{\partial x_i^{(S)}}-\frac{c_i}{D^{(S,x)}}$, where $\frac{\partial \alpha_i}{\partial x_i^{(S)}}=\frac{\partial}{\partial x_i^{(S)}}\Big( \frac{k_ix_i^{(S)}}{\sum_{j\in S}{k_jx_j^{(S)}}+k_ll} \Big)=
\frac{k_i}{\sum_{j\in S}{k_jx_j^{(S)}}+k_ll}-
\frac{{k_i}^2x_i^{(S)}}{\big(\sum_{j\in S}{k_jx_j^{(S)}}+k_ll \big)^2}$.
Thus, $\frac{\partial Z_i^{(S,x)}}{\partial x_i^{(S)}}=\frac{\beta}{D^{(S,x)}}\tau(r+\theta t_i)e^{-\beta z_i t_i }\bigg[\frac{k_i}{\sum_{j\in S}{k_jx_j^{(S)}}+k_ll}-
\frac{{k_i}^2x_i^{(S)}}{\big(\sum_{j\in S}{k_jx_j^{(S)}}+k_ll \big)^2}\bigg]-\frac{c_i}{D^{(S,x)}}$. 
Equalizing to $0$, the best response function is 
${x_i^{(S)}}^*=\frac{\big(\sum_{j\in S}{k_jx_j^{(S)}}+k_ll \big)^2}{{k_i}^2}
\bigg[ \frac{k_i}{\sum_{j\in S}{k_jx_j^{(S)}}+k_ll} - 
\frac{c_i}{\beta\tau(r+\theta t_i)e^{-\beta z_i t_i }} \bigg]=
\frac{\sum_{j\in S}{k_jx_j^{(S)}}+k_ll}{{k_i}}-
\frac{\big(\sum_{j\in S}{k_jx_j^{(S)}}+k_ll \big)^2 c_i}{{k_i}^2\beta\tau(r+\theta t_i)e^{-\beta z_i t_i }}=
\big(\sum_{j\in S}{k_jx_j^{(S)}}+k_ll\big)
\bigg[ \frac{1}{k_i}-
\frac{\big(\sum_{j\in S}{k_jx_j^{(S)}}+k_ll \big) c_i}{{k_i}^2\beta\tau(r+\theta t_i)e^{-\beta z_i t_i }}\bigg]$.
Reformulating, we have ${x_i^{(S)}}^*=
\psi^{(S)}
\bigg[\frac{1}{k_i}-
\frac{\psi^{(S)}}{{k_i}^2\beta\tau(r+\theta t_i)e^{-\beta z_i t_i }}c_i\bigg]$
where $\psi^{(S)}=\sum_{j\in S}{k_jx_j^{(S)}}+k_ll$.
By bringing $k_i$ to the left of the equality, summing on all players and adding $k_ll$ on both sides, we get 
$\sum_{j\in S}{k_jx_j^{(S)}}+k_ll=
\psi^{(S)} \bigg[ |\hat{S}|-\frac{\psi^{(S)}}{\beta\tau}
\sum_{j\in \hat{S}}{\frac{c_j}{k_j(r+\theta t_j)e^{-\beta z_j t_j}}}\bigg]
+k_ll$.
Replacing $\sum_{j\in S}{k_jx_j^{(S)}}+k_ll$ by $\psi^{(S)}$, we obtain 
$\psi^{(S)}=
\psi^{(S)} \bigg[ |\hat{S}|-\frac{\psi^{(S)}}{\beta\tau}
\sum_{j\in \hat{S}}{\frac{c_j}{k_j(r+\theta t_j)e^{-\beta z_j t_j}}}\bigg]
+k_ll$, or equivalently, 
$0=\frac{1}{\beta\tau}
\sum_{j\in \hat{S}}{\frac{c_j}{k_j(r+\theta t_j)e^{-\beta z_j t_j}}}{(\psi^{(S)})}^2-
(|\hat{S}|-1){\psi^{(S)}}-k_ll$.
Solving the equation for the positive values of $\psi^{(S)}$, we get
$\psi^{(S)}=\frac{|\hat{S}|-1+\sqrt{(|\hat{S}|-1)^2+\frac{4k_ll}{\beta\tau}
\sum_{j\in \hat{S}}{\frac{c_j}{k_j(r+\theta t_j)e^{-\beta z_j t_j}}}}}
{\frac{2}{\beta\tau}\sum_{j\in \hat{S}}{\frac{c_j}{k_j(r+\theta t_j)e^{-\beta z_j t_j}}}}$.
Therefore, we have the following equilibrium strategy:
$x_i^{(S)^*}=\max 
\bigg\{ 
\psi^{(S)}
\bigg( 
\frac{1}{k_i}-\frac{\psi^{(S)}} {{k_i}^2 \beta\tau(r+\theta t_i) e^{-\beta z_i t_i}} c_i 
\bigg),0 
\bigg\}$, where
$\psi^{(S)}=\frac{|\hat{S}|-1+\sqrt{
{(|\hat{S}|-1)}^2+\frac{4k_l l}{\beta\tau} \sum_{j\in \hat{S}}{\frac{c_j}{k_j (r+\theta t_j) e^{-\beta z_j t_j } }}}}
{\frac{2}{\beta\tau} \sum_{j\in \hat{S}}{\frac{c_j}{k_j (r+\theta t_j) e^{-\beta z_j t_j } }}}$

\end{IEEEproof}

\subsection{Values of the parameters for the simulation}

\subsubsection{General parameter $\beta$}

The constant rate of resolution of the PoW is set to $0.01$ block per minute to generate a new block on average every ten minutes. 

\subsubsection{Pool sub-game}

$r$: The current reward for mining a block that is added to the chain is $6.25$ bitcoins.
$\tau$: The value of one bitcoin is estimated to \$10,000 at the time of this work according to~\cite{Bitcoincom}.
$c_i$: The marginal cost of electricity is rounded to $0.0007$ kWh/min based on the average marginal cost of electricity estimated to $\$0.04$ kWh in~\cite{coinshares}.
$t_i$: The average number of transactions in a block is estimated to $2100$ according to~\cite{Bitcoincom} (the number of transactions per block is generally between $1500$ and $2500$).
$\theta$: The average transaction fees of $\$0.00012$ per transaction are obtained by converting in bitcoins ($1$ bitcoin = $100$M satoshis) the product of the average transaction size ($600$B) and the average charge rate per byte ($20$ satoshis/B). Likewise, transaction fees are generally between $0.00004$ and $0.0012$, values estimated from a minimum transaction size of $400$B and a maximum of $1200$B, and from a minimum fee rate of $10$ satoshis/B and a maximum of $100$ satoshis/B. All values are estimated from~\cite{Bitcoincom} and~\cite{Bitcoincom}.
$z$: Obtained from the analysis of Xiong et al.~\cite{xiong2018optimal} and set to $0.005/60$ based on a time in minutes, which corresponds to a propagation time of $10.5$ seconds for a block containing $2100$ transactions. During that time, there is a risk of a fork in the chain. This result is consistent with that of 11.37 seconds estimated in~\cite{decker2013information}.
$S$, $\hat{S}$, $\mathcal{U}$: The number of large pools that have sufficient resources to play the equilibrium strategy is estimated to $10$ based on~\cite{BlockchaincomHashDistr} on the distribution of computing power in the Bitcoin network.
$l$: Arbitrarily set to correspond to the proportion of the network computing power that does not come from large pools, which is about 1/11 of the total power according to~\cite{BlockchaincomHashDistr} and~\cite{Bitcoincom}.

\subsubsection{Protocol game}

$\lambda_i$, $\mu_i$: According to~\cite{bitnodes}, the number of nodes present in the system during a $24$ hours period varies between $10,300$ and $10,600$ approximately, with an average of around $10,450$ connected nodes. To reproduce this distribution to the miners connected to a pool, we suppose $1100$ nodes are registered in it and that $500$ of them are nonstrategic. We apply an arrival rate in the pool of $\lambda_i=1$ to the $600$ strategic miners and a departure rate of $\mu_i=0.1$. The probability density distribution obtained for the time spent in each state for a miner who connects to the system at state $S=550$ shows that almost all the time (99.97\%) is spent between states $S=520$ and $S=570$, which is consistent with~\cite{bitnodes}.

\bibliographystyle{IEEEtran}
\bibliography{bibliography.bib}

\end{document}